# Design of a Compact, Radiation Tolerant AlGaAs Geiger Photodiode

Matthew Downing[1], Erik B. Johnson[1], Joe Campbell[2], and Adam A. Dadey[2]

[1]*Radiation Monitoring Devices, Inc., 44 Hunt Street, Watertown, MA 02472*

[2]*University of Virginia, Charlottesville, VA 22904*

I. ABSTRACT

The Geiger photodiodes, or single photon avalanche photodiodes, for a radiation tolerant solid-state photomultiplier (SSPM) are being designed using Aluminum Gallium Arsenide (AlGaAs) and are less than 1 micrometer thick. Studies on the changes in dark current compared to silicon from thermal neutron damage were conducted demonstrating that the design concept is feasible, as the device sensitivity to incident radiation is low relative to silicon devices. To achieve the required high fields for Geiger mode operation within a thin device, a triple mesa structure was developed, and simulations show a favorable field profile that constrains the highest electric fields to the middle of the device and keeps the field strength at all corners and surfaces 3-6x lower than the field in the center at breakdown.

II. INTRODUCTION

Many future science discoveries will be made using experiments that feature intense radiation fields to probe more exotic features of nature. Common instruments in such experiments may include a photodetector that is coupled to a scintillation material. One photodetector, the photomultiplier tube (PMT), is known to have a high radiation tolerance, yet PMTs are bulky, draw a lot of power, and are sensitive to magnetic fields. To solve these issues with the PMT, a new type of detector called a solid-state photomultiplier was developed. A specific type of SSPM made of silicon, the silicon photomultiplier (SiPM), has the capability to provide high-resolution (<100 um) spatial resolution as position sensitive arrays of the Geiger photodiodes can be integrated into the chip [1]. The drawback for silicon is that it can readily be damaged from gamma rays and neutrons, causing the dark noise of the photodetector to increase. As an example, when the Spallation Neutron Source at Oak Ridge National Laboratory is upgraded, the neutron beams will be significantly more intense resulting in a more rapid degradation of any SiPM-based instrument compared to the existing beam conditions. To avoid the drawbacks of either PMTs or SiPMs, there is motivation for the development of SSPM detectors using non-silicon semiconductor materials, such as aluminum gallium arsenide (AlGaAs) [2-10]. This material has already shown strong potential for use in high-performance avalanche photodiodes in the 400nm to 550nm wavelength range [11-15], and GaAs is already a well-known candidate for sensor development with a wide range of applications such as solar cells on spacecraft. The use of a bandgap matched aluminum arsenide (AlAs) window on the top of a potential design could provide a means for photons to be absorbed at depth away from the top mesa surface, which can be a source of dark current post diode processing. Furthermore, a new triple-mesa geometry will help to shape the electric field and reduce the dark current.

### III. THE SILICON PHOTOMULTIPLIER

Silicon photomultipliers [16, 17] are made from an array of pixel elements that have had several names over time; these pixel elements, which can be called Geiger photodiodes (GPD), Geiger-mode avalanche photodiodes (GmAPD), or single photon avalanche diodes (SPADs), are biased beyond the breakdown voltage. When operated at this voltage (termed as Geiger-mode), a self-sustaining avalanche breakdown of electrons and holes can produce a large gain comparable to that of PMTs. Regardless of the incident photon energy or the intensity on the SPAD, the response from a single GPD is binary. As a scintillator produces a diffuse light flash, an SiPM can provide information on the light intensity from the scintillator by providing a summed signal of all the SPADs in the SiPM array that underwent an avalanche breakdown.

The radiation tolerance of SiPMs to gamma rays and neutrons has been tested previously, as discussed [18]. This reference shows that the gamma dosage was found to produce a minor increase in the dark current, with the dark current only increasing from 0.1 uA to 10 uA with a cumulative exposure of 1 Mrad. Displacement damage from neutrons in the silicon has a much more significant impact on the dark current, increasing the dark current from 0.1 uA to over 100 uA for a cumulative exposure of just 128.5 rad. The neutron exposure likely causes such a severe increase in the dark current due to the non-ionizing energy loss within the silicon from neutron scattering interactions. The dark current post-neutron damage in this reference shows a gradual recovery from "room temperature" annealing. The annealing of SiPMs has been studied in [19], where recovery back to the nominal dark current can be partially obtained and is dependent on the temperature.

A strong absorbing material, such as GaAs, allows for a thin photodiode without sacrificing the quantum efficiency. The thin device will be less sensitive to incident radiation, where the number of interactions per particle flux can be significantly less than silicon. The challenge is to provide a structure that will support a high electrical field without any deleterious field effects.

### IV. MATERIAL SELECTION

The wavelengths of interest for scintillator-based detectors range from 300nm up to 800nm, with the peak emission around 400nm. If a material's bandgap is too large, it will not be able to detect the photons generated by the scintillator; Table 1 shows the cutoff wavelengths for a variety of III-V binaries. This means that GaN and 4H-SiC can both be ruled out as they both have a bandgap (>3 eV) that is too large for electron-hole pair generation for photons with a wavelength greater than 400 nm. 3C-SiC and InN are interesting materials, but the maturity of these materials for developing new sensors is low. In addition, InN cannot be used due to the very large thermal neutron cross section of $^{115}$In.

One metric used to measure the radiation hardness of a material is the threshold displacement energy. This is the energy needed to create atomic displacement within a crystal structure and is proportional to the inverse of the lattice constant of the crystal [20], meaning a smaller lattice constant corresponds to a more radiation tolerant material. Silicon has a lattice constant of 5.43 Å. As seen in Table 1, GaAs has a similar lattice constant.

| Material | Bandgap (eV) | Cutoff (nm) | Lattice Constant (Å) | Atten. Length at 400 nm, $\alpha_{400}$ (nm) |
|---|---|---|---|---|
| GaAs | 1.424 | 871 | 5.653 | 15 |
| GaP | 2.26 | 549 | 5.45 | 116 |
| Si | 1.12 | 1107 | 5.43 | 145 |
| AlAs | 2.168 | 572 | 5.66 | 248 |
| 3C-SiC | 2.36 | 525 | 4.38 | 5121 |
| GaN | 3.4 | 344 | 4.52 | N/A |
| InN | 1.89 | 620 | 4.967 | 100 |

Table 1. Key parameters for some relevant III-V materials.

GaAs is a mature material in sensor development and has a proven track record for being more radiation tolerant compared to silicon. From [21], the threshold displacement energy for either Ga or As in GaAs is slightly smaller (9 eV) compared to silicon (13 eV), yet the advantage of GaAs is that the minimum kinetic electron (or neutron) energy to displace either the Ga or As is much higher compared to silicon. The minimum electron energy for GaAs and silicon are 342 keV to displace Ga, 372 keV to displace As, and 228 keV to displace Si, while the neutron energies are 160 keV, 179 keV, and 96 keV, respectively. Another major advantage of GaAs over silicon is that the attenuation length is ideal for the development of thin photodetectors to obtain high detection efficiencies (see Fig. 1). Improving the radiation tolerance of an AlGaAs device can be done by making it thinner, where less energy can be absorbed for a given flux of radiation. For light with a wavelength of 400nm, the attenuation length of GaAs is 15nm, and the attenuation length of AlAs is 248nm. These allow for an AlGaAs device consisting of mostly GaAs with an AlAs optical window that can be made much thinner than a silicon device, reducing the cross section for neutron capture and thus improving the radiation tolerance. Such a thin device will have large electric fields when subjected to a high bias voltage, and this must be taken into consideration when designing the device. Since it is known that the surface of a processed diode can result in high leakage current, especially when an electric field is designed to collect near surface charge, implementing a window to allow for photons to transmit into the diode and be absorbed at depth can provide a means for reducing surface leakage currents.

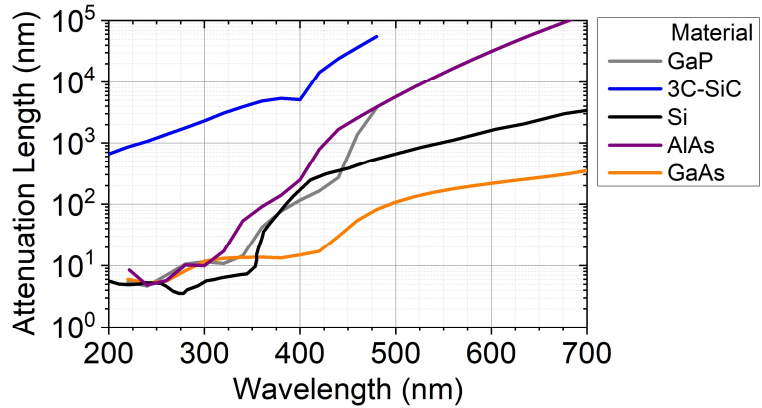

Fig. 1. The attenuation length versus the photon wavelength for various semiconductors. Color online.

## V. Radiation Studies

The development of this detector is driven by the upgrade at the Spallation Neutron Source, doubling the power of the linear accelerator from 1.4 to 2.8 megawatts and delivering a higher neutron flux to the First Target Station. With a higher neutron flux, a detector with a higher radiation tolerance is necessary to reduce how often the devices must be replaced. To test the ability of an AlGaAs device to withstand neutron dosage, the damage due to thermal neutrons on prototype AlGaAs diodes with an active thickness of approximately 600 nm was studied. At the Univ. of Massachusetts Lowell Research Reactor, there is a thermal neutron column producing neutrons with wavelength from 0.9 to 6.4 Å. Two AlGaAs diode samples and a 6-mm On-Semi J-Series SiPM were placed in the thermal neutron beam for 2.5 hours. This exposure resulted in a fluence of $10^9$ n/cm$^2$. The AlGaAs diodes are Geiger photodiodes (GPD or SPAD) by design, and the dark current post breakdown is the focus.

To account for the differences from Si diode to AlGaAs diode, we rescaled all of the biases to be set as a fraction of the breakdown voltage. The dark current from all of the AlGaAs diodes under study were summed. Once the data was rescaled and summed, the plots provide a more direct comparison between the AlGaAs and Si devices. The silicon shows a significant increase in dark current above breakdown, where the AlGaAs device does not. The SiPM active area is 27.6 mm$^2$ and the total area of the AlGaAs diodes is 1.24 mm$^2$. The fact that the AlGaAs diodes are showing no impact from the dose is a clear demonstration that it is feasible to produce a radiation tolerant AlGaAs SSPM for these environmental conditions.

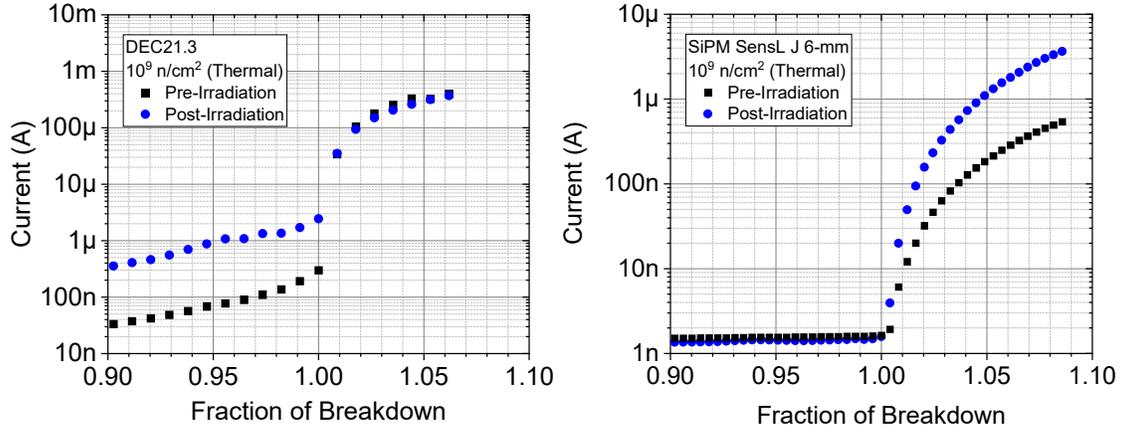

Fig. 2. Processed curves of the current versus the fraction of breakdown for the GaAs (left) and SiPM (right) before and after irradiation of cold neutrons to a fluence (dose) of $10^9$ n/cm$^2$. Color online.

## VI. THE TRIPLE MESA STRUCTURE

High-field semiconductor devices, such as Geiger photodiodes, require proper side wall termination to constrain the electric field and reduce leakage, consequentially lowering the dark current. For silicon devices, this typically takes the form of diffused regions and guard rings. The traditional means for controlling the field profile of an avalanche photodiode using a III-V semiconductor is a beveled mesa. A single-mesa structure is typically thick from top to bottom of the diode with beveled edges, which leads to a smaller electric field on the edge, as the distance from anode to cathode is longer compared to the central bulk region, where the electric field is larger. A thick device will be problematic in a high radiation environment, as it increases the likelihood of a neutron interaction and thus damage to the device.

To reduce this likelihood, a thin device was designed. To produce a thin single mesa structure that will sufficiently suppress the sidewall fields, the bevel angle would need to be very large making it difficult to etch. In order to laterally constrain the electric field, a triple mesa structure was designed. This structure splits the single mesa into three distinct sections, with the radius of the sections decreasing with each additional section. As the radius decreases it creates three mesas along the sides of the diode. These additional mesas will keep the electric field low on the edges of the device even when a high bias voltage is applied and the electric field in the bulk region is large. This effect is maintained for a wide range of thicknesses, allowing for the device to be significantly thinner, increasing the radiation tolerance while keeping the field leakage and dark current low.

## VII. TRIPLE-MESA SIMULATIONS

An AlGaAs device was simulated extensively using Lumerical, a suite of solvers for simulating photonic components, circuits, and systems. These simulations included a wide variety of device thicknesses and widths, doping concentrations, and layer layouts.

The main plot of interest generated by these simulations is the 2D electric field profile, which gives an idea of the region where breakdown multiplication will occur as well as the intensity of the field at the surface of the device.

High electric fields at the corners between the mesas in the device does lead to leakage of the field, which will be seen as an increase in the dark current of the device. The structure design was iterated in order to achieve a design that minimized these fields while also maintaining a wide avalanche multiplication region, which will provide a high probability for Geiger events to occur. The key issue is that the diode will need to be driven past the breakdown voltage, where the electric field is exceeding 4E7 V/m, while providing a low field on the edges near the device upper surface. These features are obtained in the design shown in Fig. 3 for a diode with an active thickness of approximately 745 nm. A marginal field near the surface with an AlAs window prevents near surface charge from being collected and will provide a high quantum efficiency greater than 50% for 400 nm photons and above – a quantum efficiency of 61% is estimated for 500 nm photons. These quantum efficiencies were derived from first order calculations using the index of refraction, normal incidence, and attenuation coefficients.

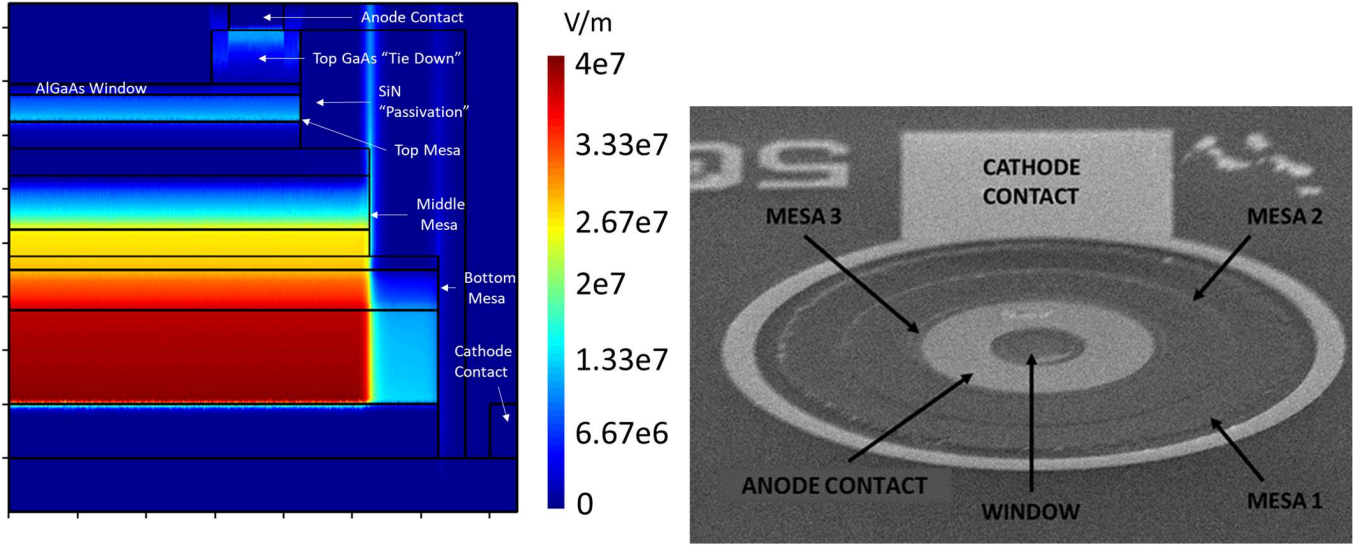

Fig. 3. Left) An AlGaAs triple mesa structure simulated in Lumerical. There is added passivation along the edge of the device. Note the high field region (red) in the bulk of the device and a low field at the edges of the device. Color online. Right) An image of the fabricated diode.

The simulated device breaks down at a bias voltage of -11V applied to the anode. However, as shown in Fig. 4, when the device is biased to -20V on the anode (an excess bias of 9V), it results in a wide breakdown region and a much more gradual decline in field strength leading up to the bottom of the AlAs window and eliminates a well that forms in the higher layers. This improves the performance of the device, as the electrons experience a greater acceleration towards the cathode, resulting in a higher number of electrons being collected.

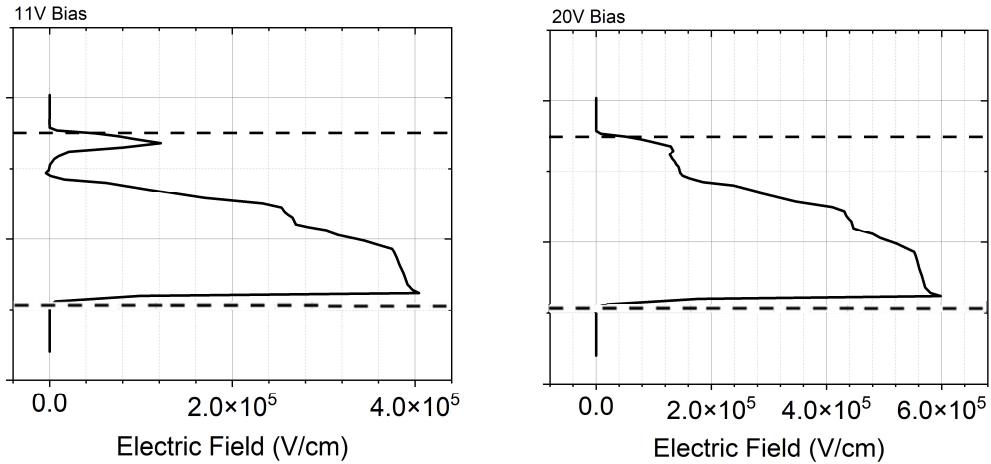

Fig. 4. The electric fields shown as a function of depth on the y-axis at a -11V bias (left) and a -20V bias (right). Both biases are applied to the anode. The dashed lines indicate the region between the n-type doping region and the bottom of the AlAs window, with the n-type doped region at the bottom of the plot.

## VIII. Results

An epitaxy wafer was grown with the desired layer thicknesses and doping concentrations. This wafer was then sent to the University of Virginia, where both single mesa and triple mesa devices were fabricated with diameters ranging from 50um to 1mm. The single mesa device was fabricated in order to test the quality of the wafer and ensure that it would function as a photodetector, and the triple mesas were fabricated from the same wafer to test how the new design affected the performance.

A current-voltage (IV) and a capacitance-voltage (CV) scan were performed on both single- and triple-mesa devices with a diameter of 200um, with a negative bias voltage applied to the anode of the device. This bias voltage was swept across the range between 0V and -11V with the current and capacitance being measured at each voltage. The IV-curve generated provides information about both the breakdown voltage and the dark current of the devices, while the CV-curve provides information about the doping concentrations throughout the device. In addition, the external quantum efficiency (EQE) of the triple-mesa device was measured.

The IV- and CV-curves for both devices are shown in Fig. 5. For both devices, breakdown occurs with a bias of -10.5V to the anode. This is very close to the breakdown voltages from the simulations, which was -11V. There is also a significant drop in the dark current for the triple mesa device compared to the single mesa device. This dark current is expected to be driven even lower if a passivation layer is applied to the device. The capacitance also drops for the triple mesa device, but the capacitance of both devices indicates that the fabricated wafer is over-doped relative to the simulations. This was confirmed by secondary ion mass spectroscopy (SIMS) analysis and will lead to a smaller breakdown region in the device while also potentially raising the dark current due to surface leakage.

The external quantum efficiency at -4 V for the triple mesa device, both with and without the central window etched, is also shown in Fig. 5. For the unetched sample, the performance of the diode exceeds expectations for having a 100nm thick layer of GaAs on top of the diode, where a 21% EQE is expected for 500 nm photons compared to the 26% found in testing. With the central window etched, the EQE for 500 nm photons increased to 52%. This is lower than the 61% expected from simulations and is likely due to the central window being under etched.

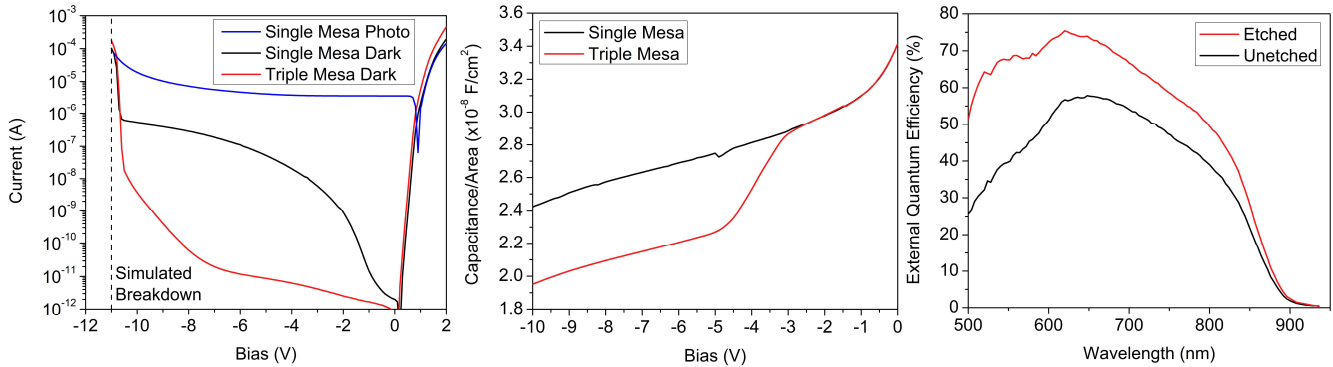

Fig. 5. Left) The IV-curves in a dark environment and under lamp illumination for both the single- and triple-mesa devices. Color online. Middle) The bottom-mesa area-normalized CV-curves for the single- and triple-mesa devices. Color online. Right) The external quantum efficiency at -4 V for the triple mesa device with and without the central window etched. Color online.

## IX. Conclusions

As the need for radiation-tolerant photodetectors rises, a solution will be needed to replace the SiPM, as it is highly susceptible to damage from radiation. A radiation tolerant SSPM is being developed using AlGaAs, a material that is already a mature material in sensor development. The optical properties of AlGaAs allow for a thin device to be made, and a triple mesa structure is implemented to obtain a high field while keeping the dark current low. Early testing on an AlGaAs diode with the triple mesa structure has given promising results, showing a high external quantum efficiency and a significantly improved dark current in the triple mesa device compared to a single mesa device, all while maintaining an active diode thickness of under 700nm. The dark current should improve even further once the diode is passivated. Once optimized, these various geometries for these AlGaAs photodiodes will be placed into arrays to form radiation tolerant SSPMs for further testing.